\documentclass[12pt,preprint]{aastex}

\shorttitle{C$_3$O towards Elias 18}
\shortauthors{Palumbo et al.}

\begin{document}

\title{Detection of C$_3$O in the low-mass protostar Elias 18}

\author{M. E. Palumbo }
\affil{INAF-Osservatorio Astrofisico di Catania, I-95123 Catania, Italy}
\email{mepalumbo@oact.inaf.it}

\author{P. Leto}
\affil{INAF-Istituto di Radioastronomia, Sez. Noto, Italy}

\author{C. Siringo}
\affil{Universit\`a degli Studi di Catania, Dipartimento di Fisica ed Astronomia, Catania, Italy}

\and

\author{C. Trigilio}
\affil{INAF-Osservatorio Astrofisico di Catania, I-95123 Catania, Italy}

\begin{abstract}
We have performed new laboratory experiments which gave us the possibility to obtain an estimate of the amount of carbon chain oxides (namely C$_3$O$_2$, C$_2$O, and C$_3$O) formed after irradiation (with 200 keV protons) of pure CO ice, at 16 K.
The analysis of laboratory data indicates  that in dense molecular clouds, when high CO depletion occurs, an amount of carbon chain oxides as high as 2-3$\times$10$^{-3}$ with respect to gas phase carbon monoxide can be formed after ion irradiation of icy grain mantles.
Then we have searched for gas phase C$_2$O and C$_3$O towards ten low-mass young stellar objects. Among these we have detected the C$_3$O line at 38486.891 MHz towards the low-mass protostar Elias 18.
On the basis of the laboratory results we suggest that in dense molecular clouds  gas phase carbon chain oxides are formed in the solid phase after cosmic ion irradiation of CO-rich icy mantles and released to the gas phase after desorption of icy mantles.
We expect that the Atacama Large Millimeter Array (ALMA), thanks to its high sensitivity and resolution, will increase the number of carbon chain oxides detected in dense molecular clouds.
\end{abstract}


\section{Introduction}
One of the main open questions in astrochemistry is the relation between solid phase and gas phase chemistry in dense molecular clouds. In fact in these regions ice mantles form on silicatic and carbonaceous grains after both direct freeze out of gas phase species and grain surface reactions.
The presence of icy grain mantles is indirectly deduced from depletion of gas phase species and directly observed in the infrared from absorption features, attributed to vibrational modes of solid phase molecules, superposed to the background stellar spectrum.
Icy grain mantles have been detected both in quiescent regions and in star forming regions. In both environments these suffer from energetic processing due to cosmic ions and UV photons. Ion and UV irradiation cause a modification of the structure and of the chemical composition of grain mantles, that is the formation of molecular species not present in the original ice. After desorption of icy mantles molecular species are released to the gas phase which could be enriched by species formed in the solid phase.

Former laboratory experiments have shown that carbon chain oxides (e.g., C$_2$O, C$_3$O, C$_3$O$_2$, C$_4$O, C$_5$O$_2$, C$_7$O$_2$)  are formed after ion irradiation and UV photolysis of CO-rich ice mixtures \citep[e.g.,][]{gsetal97, gerakines01, trottier04, loeffler05}  along with carbon dioxide (CO$_2$) which is the most abundant molecule formed \citep[e.g.,][]{gerakines96, loeffler05}.

Carbon chain oxides, namely dicarbon monoxide (C$_2$O) and tricarbon monoxide (C$_3$O) have been detected in the molecular cloud TMC-1 towards the cyanopolyyne peak (hereinafter TMC-1CP) and it has been estimated that fractional abundance of C$_2$O is about 6$\times$10$^{-11}$ and that of C$_3$O is about 1.4$\times$10$^{-10}$. These  abundances have been explained by ion-molecule gas phase reactions \citep{matthews84, brown85, ohishi91, kaifu04}.
C$_3$O has also been extensively searched for \citep{matthews84, brown85} towards other objects, some of which are rich molecular-line sources and which together encompass a wide range of physical conditions. However C$_3$O was not detected in these regions and only upper limits have been estimated.
Recently C$_3$O \citep{tenenbaum06} as well as other O-bearing species, such as H$_2$O \citep{melnick01, hasegawa06}, OH \citep{ford03}, and H$_2$CO \citep{ford04},  have been detected towards the carbon star IRC +10216. This is an asymptotic giant branch star. O-bearing molecules are not expected in carbon stars since the bulk of available oxygen is contained in CO. In order to explain these observations it has been suggested that the increase in the star luminosity is causing the evaporation of orbiting icy bodies \citep{melnick01}.   Alternatively it has been suggested that gas-phase oxygen-rich chemistry is occurring in the outer shell of the star \citep{tenenbaum06}.

Here we discuss new laboratory experiments which confirm the formation of carbon chain oxides after ion irradiation of CO ice at low temperature and give us the possibility to obtain a quantitative estimate of the amount of carbon chain oxides formed with respect to initial carbon monoxide (Section 2). We present the new detection of the C$_3$O line at 38486.891 MHz towards the low-mass protostar Elias 18 and we confirm the detection of the same line towards TMC-1CP as already reported by \citet{kaifu04} (Section 3). On the basis of our laboratory results we suggest that gas phase carbon chain oxides in dense molecular clouds are actually formed in the solid phase after ion irradiation of CO-rich icy grain mantles and released to the gas phase after desorption of icy mantles (Section 4).

\section{Experimental results}
\begin{figure}
\plotone{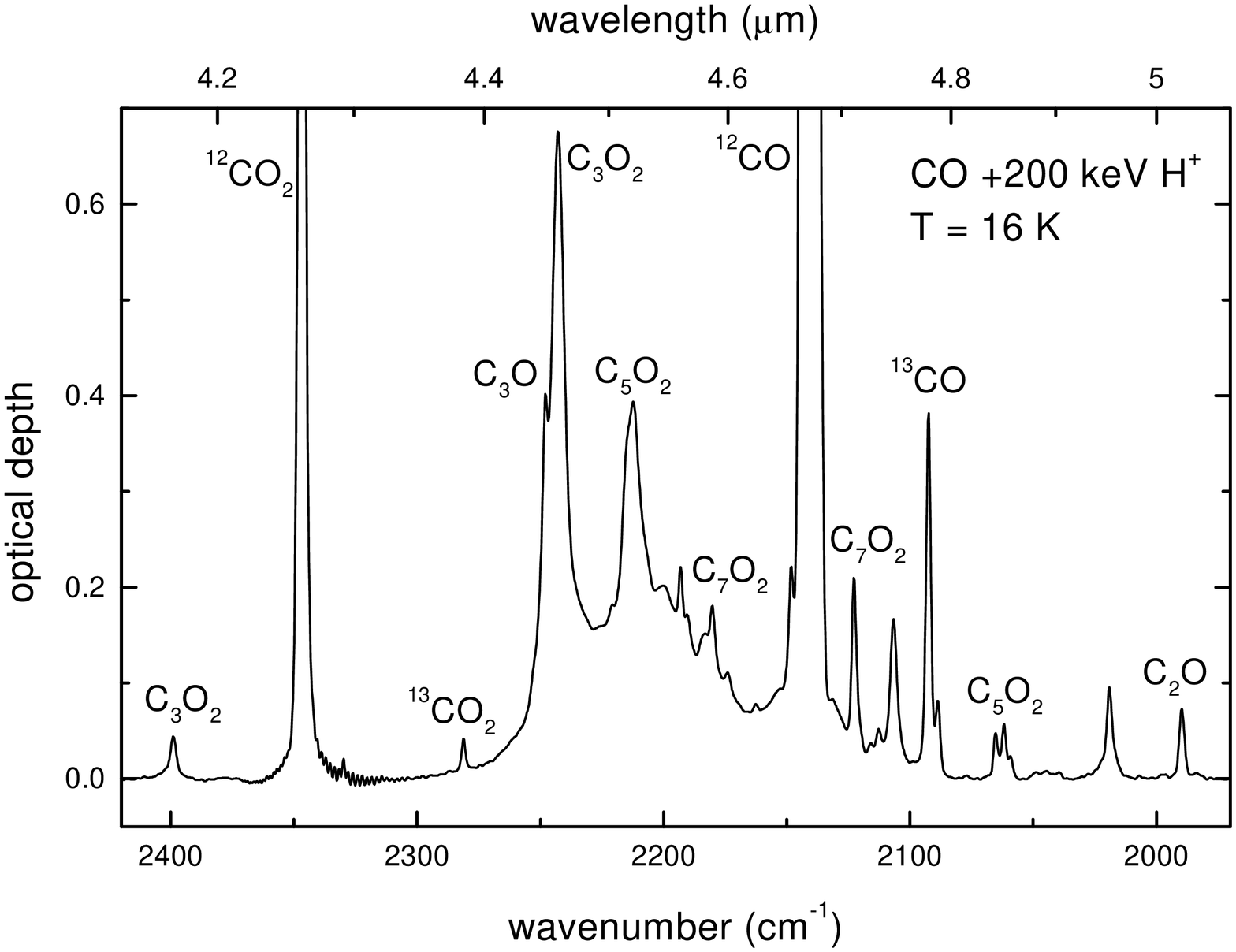}
\caption{Infrared transmission spectrum, in optical depth scale, of CO ice after ion irradiation with 200 keV H$^+$ (fluence = 1.25$\times$10$^{14}$ ions cm$^{-2}$) at 16 K. Labels indicate main species present after irradiation.}
\label{coirr}
\end{figure}

Experiments have been performed in the Laboratory for Experimental Astrophysics (INAF - Catania Astrophysical Observatory, Italy). A detailed description of the experimental set up can be found elsewhere \citep[e.g.,][]{gsetal01, ionuv02, palumbo04}. CO icy samples have been prepared on a cold substrate (16 K) in a vacuum chamber (P=10$^{-7}$ mbar) and then irradiated with 200 keV H$^+$ ions. Mid-IR transmission spectra (2.5-25 $\mu$m; 4000-400 cm$^{-1}$) have been taken before, during and after irradiation. Different CO icy samples have been irradiated and the highest ion fluence was 1.5$\times$10$^{15}$ ions cm$^{-2}$.
Figure~\ref{coirr}  shows the spectrum of pure carbon monoxide in the 2450-1950 cm$^{-1}$ (4.08-5.1 $\mu$m) spectral range at 16 K after irradiation with 200 keV H$^+$ ions (fluence equal to 1.25$\times$10$^{14}$ ions cm$^{-2}$). It is evident that after irradiation several  absorption features are present which indicate the formation of molecular species not present in the original ice sample.  In Figure~\ref{coirr}  the intense CO band at about 2139 cm$^{-1}$ (4.67 $\mu$m) and the CO$_2$ band at about 2343 cm$^{-1}$ (4.27 $\mu$m) are out of scale. Main carbon chain oxides bands formed after irradiation are labeled.
In Table~\ref{bands} the  peak position of most intense bands is reported along with their identification which is based also on the results presented by \citet{trottier04}. A few features listed in Table~\ref{bands}  still remain unidentified and further laboratory experiments will be necessary for their identification.

\begin{figure}
\plotone{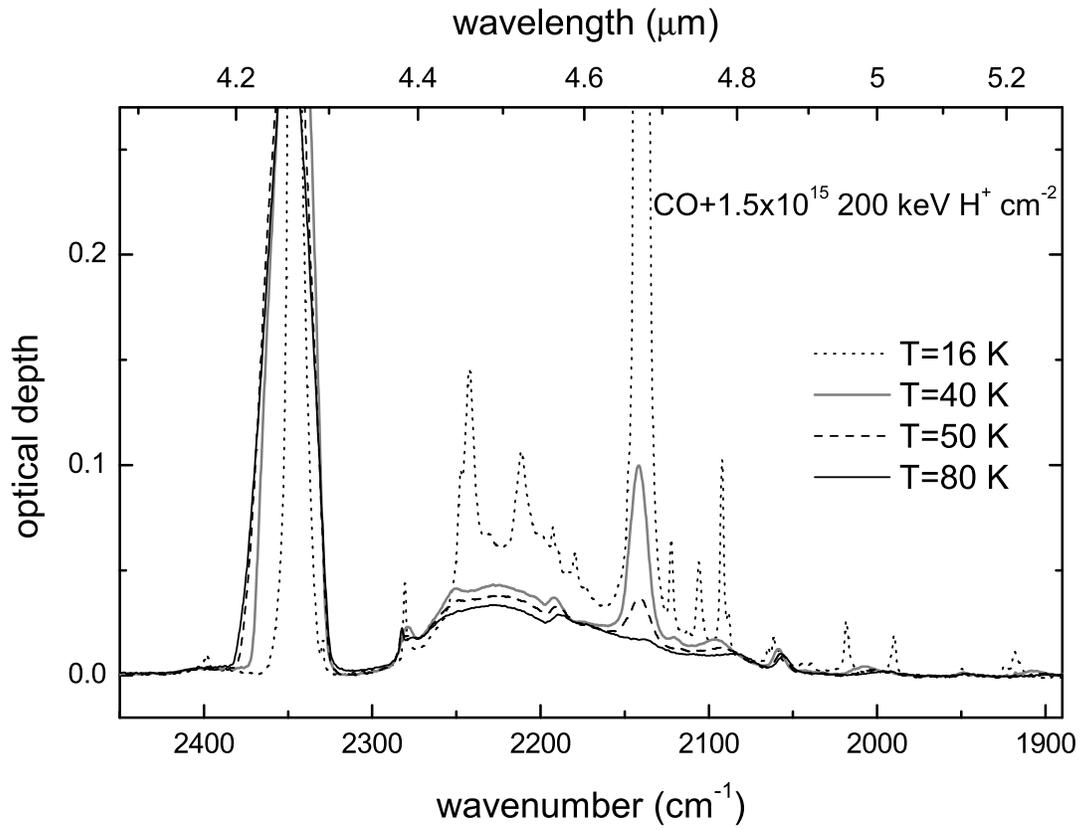}
\caption{Infrared transmission spectra, in optical depth scale, of CO ice after ion irradiation with 200 keV H$^+$ (fluence = 1.5$\times$10$^{15}$ ions cm$^{-2}$). Spectra taken at 16 K and after warm up are shown.}
\label{sublimation}
\end{figure}

Figure~\ref{ratios}  shows the column density of carbon suboxide (C$_3$O$_2$), tricarbon monoxide (C$_3$O), and dicarbon monoxide (C$_2$O) with respect to initial carbon monoxide as a function of ion fluence after irradiation of CO ice at 16 K. Column density of C$_3$O$_2$ has been obtained from the band at 2398 cm$^{-1}$  using the integrated absorbance value  of 0.8$\times$10$^{-17}$ cm molecule$^{-1}$ given by \citet{gerakines01}. The column density of C$_3$O and C$_2$O have been obtained from the bands at 2247 cm$^{-1}$ and 1989 cm$^{-1}$ respectively. We note that the 2247 cm$^{-1}$ band is superposed to the broader band at about 2242 cm$^{-1}$ due to C$_3$O$_2$. In order to estimate the integrated intensity (area) of this band we have considered a linear baseline between 2246.7 and 2249.6 cm$^{-1}$. Given that the integrated absorbance of bands due to C$_3$O and C$_2$O has not been measured we have assumed a  value of 1$\times$10$^{-17}$ cm molecule$^{-1}$ for both bands. In fact this value is close to the average value measured for the absorption bands of many molecules \citep[e.g.,][]{mulas98, kerkhof99, bennett04}.
The initial column density of carbon monoxide (the 2140 cm$^{-1}$ band being saturated) has been measured from the interference curve given by a He-Ne laser beam reflected both by the vacuum-film and film-substrate interfaces \citep[e.g.,][]{josa98} during accretion of the ice sample. An ice density equal to 0.8 g cm$^{-3}$ \citep{loeffler05} was used to obtain the column density value.
The ratio of the column density of each species relative to the initial CO column density as a function of ion fluence  has been fitted with the exponential curve
$ y = A (1-e^{-\sigma\Phi})$
where $\Phi$ is the ion fluence (ions cm$^{-2}$), $\sigma$ is the process cross section (cm$^2$) and A is the asymptotic value of the column density ratio. The $\sigma$ and A values obtained for each species are reported in Fig.~\ref{ratios}. In the case of C$_3$O$_2$ only data points obtained at ion fluence lower than 1.25$\times$10$^{14}$ ions cm$^{-2}$ have been considered in the fitting procedure. In fact we noted that at higher  fluence the column density of carbon suboxide decreases indicating that this species is destroyed after further irradiation. This also occurs to dicarbon monoxide and tricarbon monoxide at fluences higher than those given in Fig.~\ref{ratios}. This effect has in fact already been observed for species formed after ion irradiation of other ice samples \citep[e.g.,][]{ionuv02}. As discussed in details in Section 4, the top x-axis in Fig.~\ref{ratios} gives an estimation of the time (years) necessary to obtain on interstellar ices the same effects observed in laboratory.
After irradiation was completed ice samples have been warmed up and spectra have been taken at 25, 40, 50, 60, 70 and 80 K. The spectrum at 25 K is almost indistinguishable from that taken at 16 K.  Spectra taken at 16, 40, 50 and 80 K are shown in Fig.~\ref{sublimation}. It is evident that the intensity of absorption bands due to CO and carbon chain oxides rapidly decreases after warm up and that these are not detectable at about 80 K. On the other hand the bands due to carbon dioxide are present in the spectra at higher temperatures. This indicates that volatile species such as CO and carbon chain oxides sublimate.
Finally it is worth mentioning that carbon chain oxides are not formed after ion irradiation of O-rich ice samples such as CO$_2$:CO and H$_2$O:CO mixtures \citep{gsetal97}.

\section{Observations}

Radio observations have been performed in 2006 September 6-12 with the 32-m radiotelescope in Noto (Italy). Sources observed are listed in  Table~\ref{objects}.
The telescope has an active surface system that compensates the gravitational deformation of the primary mirror making  observations at high frequencies also possible. This new antenna set up, together with the favorable weather conditions, allow to operate with good performance at high frequencies.

The observations have been carried out using the 22 GHz and the new 43 GHz (that can be tuned in the 38-47 GHz range) receivers.
The beam size of the telescope (HPBW) was about $115\arcsec$ at 22 GHz and $54\arcsec$ at 43 GHz,
the system noise temperature (including atmospheric noise and antenna ohmic losses) is about 90-120 K at zenith depending on weather conditions.
The aperture efficiency of the telescope was about 0.28 at 43 GHz and 0.44 at 22 GHz.
A duty cycle of 5 min integration on source and 5 min off source was used.
Total integration time for each source is reported in Table~\ref{log}.

The spectra have been acquired with the ARCOS autocorrelator \citep{comoretto90} in beam switch mode;
two spectra, one for each polarization, have been acquired simultaneously. Each spectrum has a bandwidth of 20 MHz,
with a spectral resolution of 37 kHz, corresponding to a velocity resolution of 0.27 and 0.53 km s$^{-1}$ at 43 and 22 GHz respectively.
The determination of the antenna efficiency as a function of the elevation has been performed by observing the flux calibrators 3C286, 3C123 and NGC7027.
The spectra were reduced by using the software XSpettro for the on-off difference then software CLASS for the baseline subtraction and temperature calibration.
The antenna temperature has been corrected to account for the variation of the effective area of the telescope as a function of elevation. Then all spectra have been averaged.

\begin{figure}
\plotone{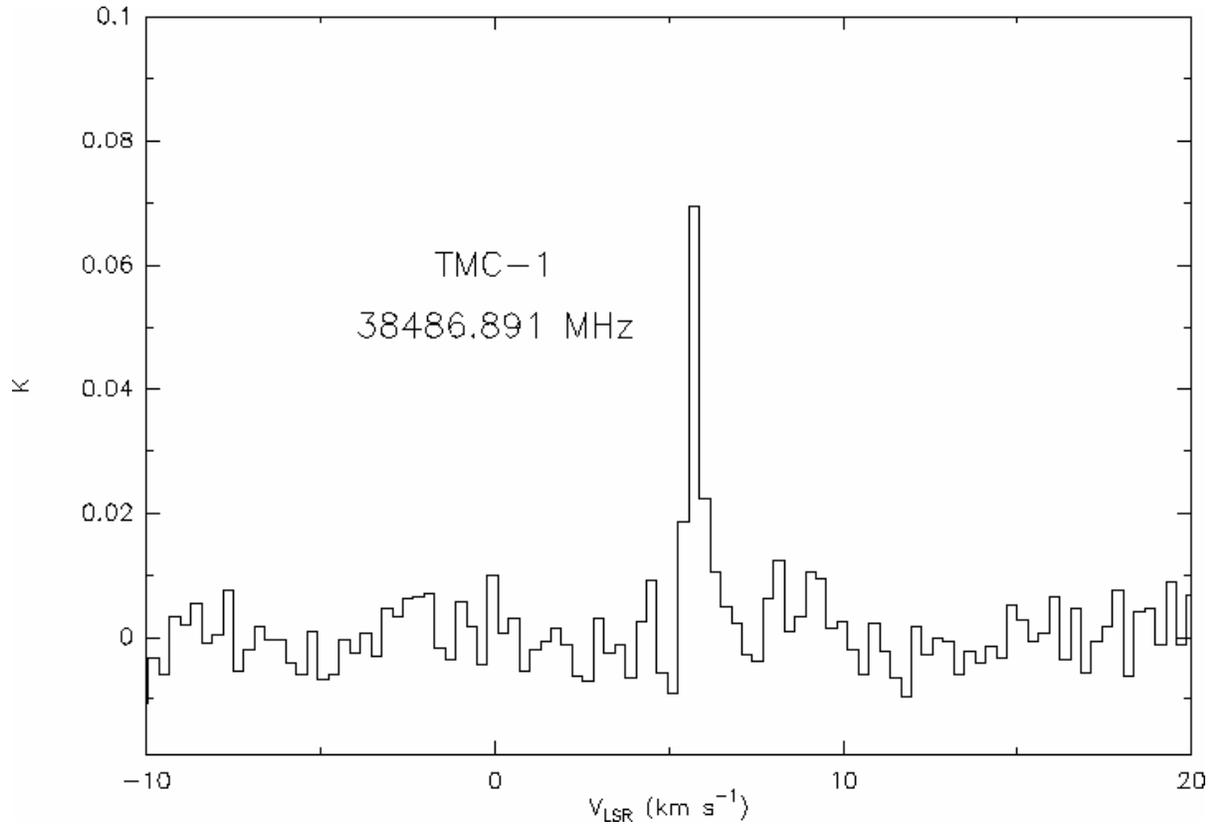}
\caption{C$_3$O line at 38486.891 MHz detected towards TMC-1CP. }
\label{tmc1}
\end{figure}

\begin{figure}
\plotone{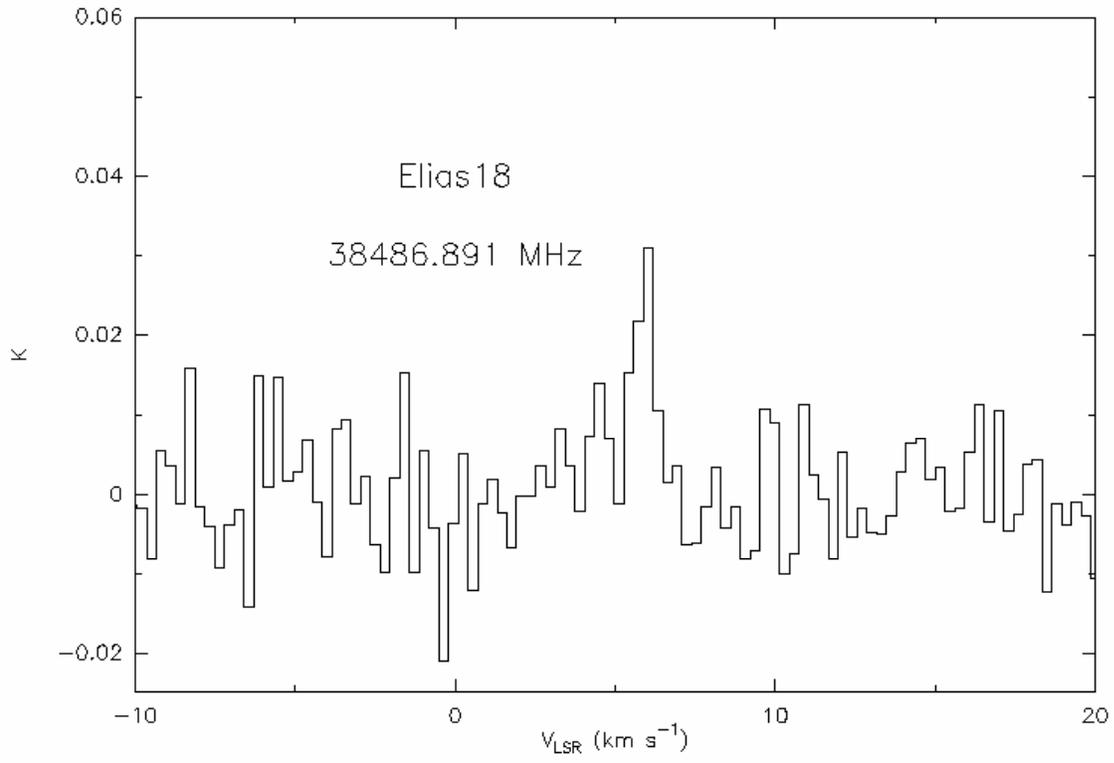}
\caption{C$_3$O line at 38486.891 MHz detected towards Elias 18.}
\label{elias18}
\end{figure}

We have searched for the C$_2$O lines at 22258.181 MHz and 45826.706 MHz and  the C$_3$O line at 38486.891 MHz towards the  low-mass young stellar objects  listed in Table~\ref{objects}. Our sample also includes TMC-1CP where the above lines have already been detected \citep{matthews84, brown85, ohishi91, kaifu04}.
Figure~\ref{tmc1}  shows the C$_3$O line at 38486.891 MHz towards TMC-1CP. This detection confirms the previous results. We have not been able to confirm the detection of C$_2$O towards TMC-1CP as reported by \citet{ohishi91}  and \citet{kaifu04}  probably due to the low S/N ratio of our measurements.
Figure~\ref{elias18}  shows the C$_3$O line at 38486.891 MHz towards Elias 18. The peak intensity of the line is 0.031 K (S/N=4.4 with rms = 0.007 K). Rms for all the sources of our sample are given in Table~\ref{log}.
Fitting the line profile with a gaussian curve, we have measured that the integrated intensity of the line observed towards TMC-1CP  is 0.039$\pm$0.008 K km s$^{-1}$ and towards Elias 18 is 0.025$\pm$0.005 K km s$^{-1}$. In the hypothesis that the source fills the antenna beam, assuming that the line is optically thin, using T$_{ex}$=20 K, $\mu$(C$_3$O)=2.39 D, and B$_0$=4810.885 MHz \citep{brown83}, following the procedure described by \cite{goldsmith99}, we have estimated that C$_3$O column density is 8$\pm$2$\times$10$^{11}$ molecules cm$^{-2}$ towards TMC-1CP and 5.2$\pm$1.3$\times$10$^{11}$ molecules cm$^{-2}$ towards Elias 18.
The value of the C$_3$O column density towards TMC-1CP we have found is lower than the value (1.4$\pm$0.4$\times$10$^{12}$ molecules cm$^{-2}$) reported by \citet{matthews84} and \citet{brown85}, however we believe that within the uncertainties of the estimation these values are comparable.
Assuming N(H$_2$)$\simeq$1$\times$10$^{22}$ molecules cm$^{-2}$ \citep{guelin82}, we obtain an abundance of C$_3$O w.r.t. hydrogen of about 10$^{-11}$.

\section{Discussion}
\begin{figure}
\plotone{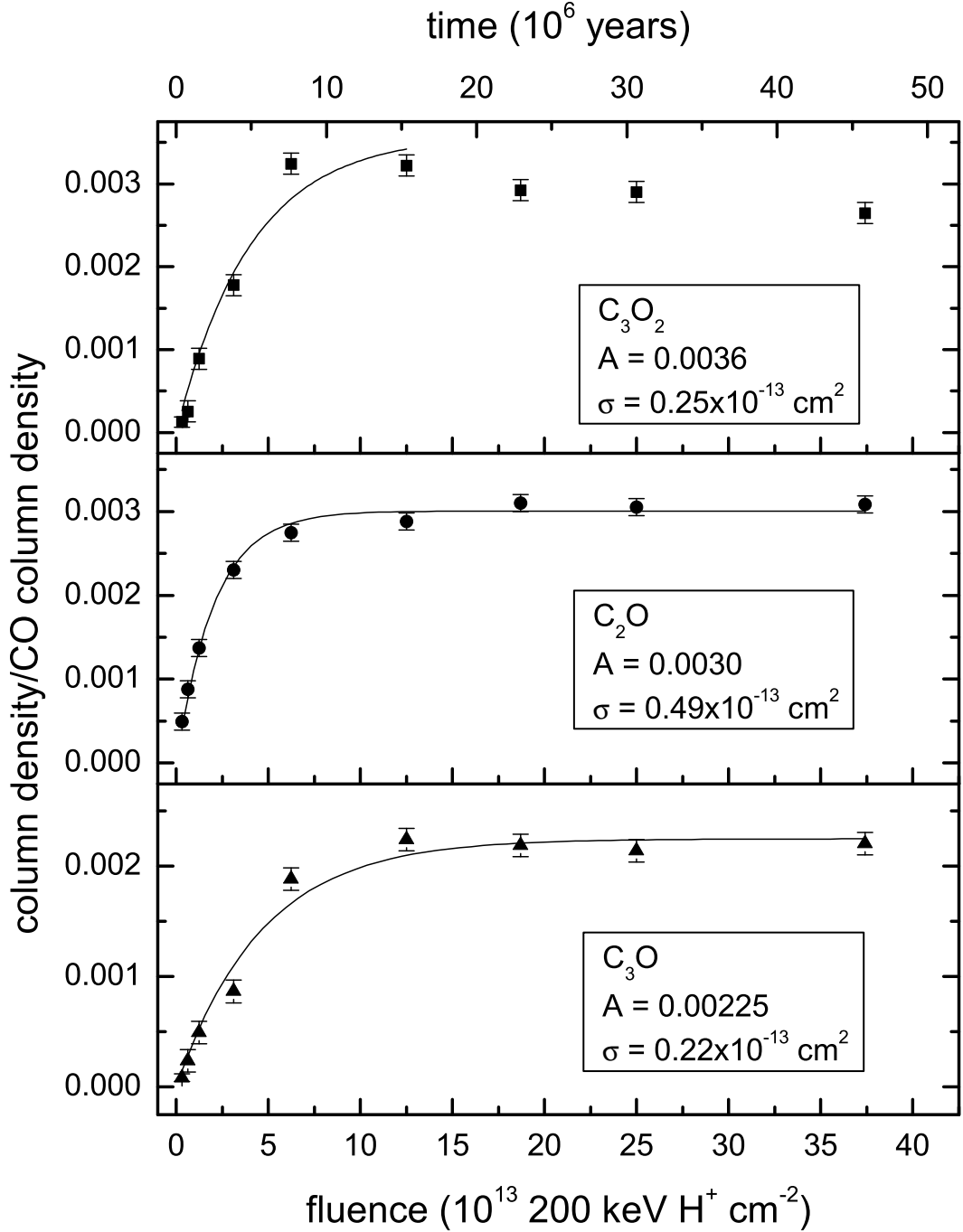}
\caption{Column densities of C$_3$O$_2$, C$_2$O and C$_3$O with respect to column density of initial CO as a function of fluence after ion irradiation of pure CO at 16 K with 200 keV H$^+$. Top x-axis indicate the time (years) which would be necessary to obtain in dense molecular clouds the same effects observed in laboratory. The experimental data have been fitted with an exponential curve (solid lines) and the fit parameters for each species are reported.}
\label{ratios}
\end{figure}

Observations have shown that in dense molecular clouds the fractional abundance of carbon chain oxides (namely C$_2$O and C$_3$O) is of the order of 10$^{-11}$-10$^{-10}$ \citep[][this work]{matthews84, brown85, ohishi91, kaifu04}. In these regions the fractional abundance of CO is of the order of 10$^{-4}$ then the abundance of carbon chain oxides with respect to CO values about 10$^{-7}$-10$^{-6}$.
Laboratory experiments here presented indicate that after ion irradiation of pure CO ice at 16 K the amount of carbon chain oxides formed is of the order of 2-3$\times$10$^{-3}$ with respect to initial CO (Figure~\ref{ratios}).
In order to estimate the time necessary to obtain in dense molecular clouds the  effects observed in the laboratory  we consider the approximation of effective monoenergetic 1 MeV protons and assume that in dense interstellar regions the 1 MeV proton flux is equal to 1 proton cm$^{-2}$ s$^{-1}$ (see \citet{mennella03} for a detailed discussion).  However our experimental results were obtained using 200 keV protons. Thus in order to extrapolate the laboratory results to the interstellar medium conditions we assume that they scale with the stopping power (S, energy loss per unit path length) of impinging ions. Using the TRIM code \citep{ziegler03} we have estimated that for protons S(200keV)/S(1MeV) is 3.8 in the case of pure CO ice. With these hypotheses in mind we have indicated in Fig.~\ref{ratios}  timescale axis (top x-axis), which gives an estimation of the time (years) necessary to obtain the  effects, observed in laboratory, on interstellar ices. Thus if we assume high CO depletion and that the carbon chain oxides/CO column density ratio obtained in the solid phase after ion irradiation is maintained in the gas phase after desorption of icy grain mantles, then from the exponential equation used to fit the data, we obtain that about 10$^3$ years would be necessary to form the observed column density of carbon chain oxides. As we will discuss below, this time is much shorter than the evolution time scale of dense clouds, thus the observed gas phase abundance of carbon chain oxides could be easily reached even if carbon monoxide is not completely depleted and/or only partial desorption of icy grain mantles takes place.

Towards all the young stellar objects  observed in this work, the solid CO absorption band at 4.67 $\mu$m has been detected \citep[e.g.,][]{kerr93, chiar94, chiar95, teresa98} along with the 3 $\mu$m band due to water ice which is the most abundant solid phase species along these lines of sight.
However, recently, a detailed study of the solid CO band profile observed towards a large sample of low mass embedded objects \citep{ponto03} has shown that typical lines of sight have 60-90\% of the solid CO in a pure or nearly pure form, suggesting that  interstellar ices are best represented by a layered model rather than a mixed ice \citep[e.g.,][]{fraseretal04}.
The presence of solid CO towards TMC-1CP has never been reported. In Taurus Molecular Cloud a threshold extinction  A$_{th} \sim 6$ mag is required for CO ice detection \citep{chiar95}. Gas-phase models use  A$_V$=10 mag in TMC-1 cores \citep[e.g.,][]{parketal06} while A$_V$=32 mag towards TMC-1A \citep{teixeira99} thus it seems reasonable to assume that CO ice is also present along the line of sight of TMC-1CP.

Elias 18 resides in a part of the Taurus molecular cloud known as Heiles cloud 2 (HCL2) where low-mass star formation is active.
It is a highly obscured object (A$_V$ $\sim$ 15-19 mag) and it has been suggested \citep{tegler95} that it is in transition between an embedded young stellar object and an exposed T Tauri star. The IR spectral energy distribution (SED) for this source is typical of a class II or ``flat-spectrum'' YSO with significant optical extinction \citep{elias78}. Recent observations indicate that Elias 18 has a circumstellar disk oriented close to edge-on and that most of the CO in the disk is incorporated in icy mantles on dust grains, i.e. high depletion is observed \citep{shuping01}.
Mid-IR observations towards Elias 18 show the presence of both the solid CO and CO$_2$ absorption bands \citep[e.g.,][]{tielens91, chiar95, nummelin01}. The comparison between observations relative to solid CO and laboratory spectra indicates that different ice mixtures can equally well reproduce the observed band profile \citep{palustraz93, chiar95, chiar98}. However all the fits indicate that a comparable amount of solid CO is in the nonpolar (i.e., CO-rich) and polar (i.e., H$_2$O-rich) components. Among the fits obtained,  it has been shown that the nonpolar component  can be reproduced by the spectrum of ion irradiated pure CO ice (and this is compatible with the hypothesis that CO-rich icy mantles are present along the line of sight in order to form  carbon chain oxides) and the polar component can be reproduced by the spectrum of CO formed after ion irradiation of a H$_2$O:CH$_3$OH ice mixture \citep{palustraz93}.
Detection of the stretching mode band of solid CO$_2$ towards Elias 18 and the comparison of the observed band profile with laboratory spectra have been reported by \citet{nummelin01}.  However, as discussed by \citet{ehrenfreund97, gerakines99, ioppolo08}, the profile of the stretching mode band does not strongly depends on the ice mixture and cannot be used to constrain the ice composition along the line of sight.

TMC-1CP is a dense core in the TMC-1 cold dark cloud. Based on gas phase observations and chemical evolution models it has been deduced that its age is about  10$^5$~years  and the density is n$_H$ = 2$\times$10$^{4}$ cm$^{-3}$ \citep{parketal06}. Thus the gas would take 10$^9$/n$_H$=5$\times$10$^{4}$ years to condense on grains \citep{xander87}. The presence of gas phase species  implies that  desorption processes, such as photodesorption, grain-grain collisions, cosmic ray induced desorption and turbulence, compete with mantle accretion in this region \citep{boland82, hasegawa93, bringa04}.

Thus detection of C$_3$O towards these lines of sight  is compatible with the hypothesis that this molecular species is formed in the solid phase and released to the gas phase when desorption of icy mantles takes place.

The results here discussed do not exclude that carbon chain oxides are also formed after gas phase reactions in dense molecular clouds.
Furthermore we are aware that further observational data are necessary to confirm these results. In fact we plan to search for carbon chain oxides towards other sources in particular hot corinos in Class 0 low mass protostellar objects where evidence of ice mantle evaporation has been reported \citep{bottinelli07}.

One of the mysteries of interstellar chemistry is the mechanism regulating the balance between gas phase and solid phase species. Carbon chain oxides could be key molecules in this field and thanks to its high sensitivity and resolution the Atacama Large Millimeter Array (ALMA) will give important results increasing the number of detected features in a larger sample of molecular clouds.

Finally, as far as we know, the detection of C$_2$O and C$_3$O  in comets has never been reported. However comets suffer from heavy ion irradiation \citep{gsetbob91} and CO is abundant in these objects thus we expect  that carbon chain oxides are present in comets too. In fact a tentative detection of carbon suboxide (C$_3$O$_2$) in comet Halley has been reported \citep{huntress91, crovisier91}. A firm  detection of carbon chain oxides  in comets could also be used to support the hypothesis that the presence of O-bearing species, and in particular C$_3$O, in carbon stars, such as IRC +10216, is due to sublimation of orbiting icy bodies as suggested by \citet{melnick01}.

\acknowledgments
We would like to thank F. Spinella for his technical support during laboratory experiments, C. Contavalle and C. Nocita for their assistance during observations at Noto. This research has been financially supported by INAF and MIUR research contracts.

\clearpage

\clearpage

\begin{table}
\caption{Most intense features detected after  irradiation with 200 keV H$^+$ ions of CO  ice at 16 K  and their identification. \label{bands}}
\begin{center}
\begin{tabular}{cc}
\tableline\tableline
band position &  identification \\
(cm$^{-1})$  & \\
\tableline
3706 &  CO$_2$ \\
3601 &  CO$_2$ \\
3069 &  C$_3$O$_2$\\
2398 &  C$_3$O$_2$\\
2340 &  CO$_2$\\
2329 &  C$^{16}$O$^{18}$O\\
2280 &  $^{13}$CO$_2$\\
2247 &  C$_3$O\\
2242 &  C$_3$O$_2$\\
2211 &  C$_5$O$_2$ \\
2192 &  OCC$^{13}$CO\\
2179 &  C$_7$O$_2$\\
2140 &  CO \\
2122 &  C$_7$O$_2$\\
2112 &  C$^{17}$O\\
2105 &  \\
2092 &  $^{13}$CO\\
2088 &  C$^{18}$O\\
2064 &  \\
2061 &  C$_5$O$_2$ \\
2018 &  \\
1989 &  C$_2$O \\
1924 &  \\
1918 &  C$_4$O\\
1042 &  O$_3$\\
\tableline
\end{tabular}
\end{center}
\end{table}

\clearpage

\begin{table}
\caption{Low-mass young stellar objects observed. \label{objects}}
\begin{tabular}{lcccc}
\tableline\tableline
Object\tablenotemark{a} & Region & $\alpha$(2000) & $\delta$(2000) \\
\tableline
TMC-1CP & Taurus &  04 41 45.9 & 25 41 27  \\
TMC1-A & Taurus & 04 39 35.0 & 25 41 47  \\
Elias 18 & Taurus & 04 39 55.7 & 25 45 02   \\
L1551 IRS5 & Taurus & 04 31 33.9 & 18 08 08 \\
L1489IR & Taurus & 04 04 43.1 & 26 18 58   \\
Elias 29 (WL15)& Ophiuchus & 16 27 09.3 & -24 37 21  \\
Elias 32 (VS18)& Ophiuchus & 16 27 28.5 & -24 27 17  \\
WL5 & Ophiuchus & 16 27 18.0 &  -24 28 52   \\
CK1 (SVS20) & Serpens & 18 29 57.5 & 01 14 07 \\
SVS4 & Serpens &  18 29 57.8 & 01 12 48  \\
\tableline
\end{tabular}
\tablenotetext{a} {Other ID is given in brackets.}
\end{table}

\clearpage

\begin{table}
\caption{Observation log and rms measured.\label{log}}
\begin{tabular}{lcccccc}
\tableline\tableline
 & \multicolumn{2}{c}{CCO }& \multicolumn{2}{c}{CCO }& \multicolumn{2}{c}{C$_3$O}\\
 &\multicolumn{2}{c}{$\nu$=22258.181 MHz}&\multicolumn{2}{c}{$\nu$=45826.706 MHz}&\multicolumn{2}{c}{$\nu$=38486.891 MHz} \\
&\multicolumn{2}{c}{J$_N$=2$_1$-1$_0$}&\multicolumn{2}{c}{J$_N$=3$_2$-2$_1$}& \multicolumn{2}{c}{J=4-3} \\
Source  & int. time &   rms  &   int. time &   rms &    int. time &   rms  \\
        &    (min)  &   (K)  &   (min)   &    (K)&     (min)   &    (K) \\
\tableline
TMC-1CP &    300    &   0.008   &  140      &   0.011   &      380    &      0.005  \\
TMC-1A &           &           &           &           &      160    &      0.010   \\
Elias 18 &  180    &   0.010   &  140      &   0.008 &      200    &      0.007  \\
L1551 IRS5 &    300    &   0.007   &  150      &    0.009  &      300    &      0.007   \\
L1489IR &   300     & 0.008     &  220      &    0.001 &    240    &    0.008    \\
Elias 29 &  90 &    0.016  &   90    &        0.019 &     200  &     0.008  \\
Elias 32 &  90 &      0.016&    90   &       0.022    &     160  &     0.011   \\
WL5     &   90 &      0.018&    80   &     0.025      &     160  &     0.009 \\
& & & & & &  \\
CK1  &      120 &    0.013 &   120 &         0.014   &    230   &      0.006    \\
SVS4-4 &    200 &   0.011  &   120 &        0.015    &      160 &      0.008   \\
\tableline
\end{tabular}
\end{table}

\end{document}